\documentclass[aps,twocolumn,epsfig,showpacs]{revtex4-1} 

\usepackage{amsmath} 
\usepackage{amssymb} 
\usepackage{graphicx}
\usepackage{color}

\begin{document}

\title{Ratchet Effect Driven by Coulomb Friction: the Asymmetric Rayleigh Piston}
 
\author{A. Sarracino, A. Gnoli, and A. Puglisi}
\affiliation{CNR-ISC and Dipartimento di Fisica, Universit\`a Sapienza, p.le A. Moro 2, 00185 Roma, Italy}

\begin{abstract}
The effect of Coulomb friction is studied in the framework of
collisional ratchets. It turns out that the average drift of these
devices can be expressed as the combination of a term related to the
lack of equipartition between the probe and the surrounding bath, and
a term featuring the average frictional force. We illustrate this
general result in the asymmetric Rayleigh piston, showing how Coulomb
friction can induce a ratchet effect in a Brownian particle in contact
with an equilibrium bath. An explicit analytical expression for the
average velocity of the piston is obtained in the rare collision
limit. Numerical simulations support the analytical findings.
\end{abstract}

\pacs{05.40.-a, 02.50.Ey, 05.20.Dd}

\maketitle

\emph{Introduction}.-- The problem of extracting work from unbiased
noise is a central issue in the context of energy harvesting at small
scales~\cite{gamma,CMAB11,MCB13}.  The phenomenon of the rectification
of non-equilibrium fluctuations is called ``ratchet effect'', from the
seminal works of Smoluchowski~\cite{s12} and Feynman~\cite{f63}, and
is studied in the theory of Brownian motors~\cite{A97,R02,HM03}. This
phenomenon can be achieved in the presence of dissipation, i.e. under
statistical non-equilibrium conditions. It requires that both temporal
and spatial symmetries are broken.  Whereas these two fundamental
constraints have been pointed out~\cite{f63}, the specific mechanisms
ruling the action of ratchet devices in their different realizations
still deserve a thorough study within general
theories~\cite{seifert,FKS12}.
 
A class of ratchet models is represented by the ``collisional
ratchets'', where fluctuations are induced by the interaction of an
asymmetric probe with one (or more) gas(es) of particles. In these
models, realized in several
variants~\cite{meurs1,meurs,vandenbroeck,cleuren2,costantini1,costantini2,cleuren},
dissipation is introduced in two ways: i) the probe is in contact with
several baths at different temperatures; ii) interactions between the
bath particles and the probe are dissipative (granular ratchets). The
rectification of thermal fluctuations has also been observed in
numerical simulations of an asymmetric particle diffusing in a glass
forming liquid~\cite{GSVGP10}. In this case, the energy flux
sustaining the motor effect is induced by the thermal unbalance
between fast and slow degrees of freedom.

Here we focus our attention on a different source of dissipation,
introduced by the presence of Coulomb friction affecting the dynamics
of the probe (or tracer) between successive collisions. This kind of
dissipation modifies the dynamics of the
tracer~\cite{dGen05,DCdG05,H05,BTC11}, and also plays an important
role in the context of ratchet
models~\cite{BBdeG06,FKPU07,talbot2,talbot1,BS12,gnoli}. Here we bring
to the fore the structural elements common to general collisional
ratchets when several forms of dissipation are present. We consider a
general stochastic model ruling the dynamics of the probe's velocity,
where the interactions with the gas particles are described by a
Master Equation and the presence of Coulomb friction is introduced
through a deterministic force. For this model we derive an expression
for the average drift where two terms appear: the first one takes into
account the lack of equipartition due to non-equilibrium conditions,
while the second one is directly related to the frictional force.  The
first term represents the heat flux exchanged by the probe with the
particles bath: it can be induced by a non-equilibrium coupling with
the bath (e.g. if dissipative interactions or \emph{reservoirs} at
different temperatures are present), and/or by frictional dissipation.
 
To illustrate this result, we study the asymmetric Rayleigh
piston~\cite{macdonald} in the presence of Coulomb friction. In this
model a Brownian tracer interacts with two gases of particles with
different masses but at the same temperature.  We show that the
dissipation through friction, coupled with the spatial asymmetry
introduced by different masses of gas molecules, is sufficient to
induce a net drift on the piston, even if the gases are in equilibrium
at the same temperature. We stress that the motor effect observed in
our model originates from a mechanism different from those
acting in systems where the piston is in contact with gases at
different temperatures~\cite{GP99,meurs}, or with gas particles with
different restitution coefficients~\cite{costantini2,talbot1}: in all
these systems, indeed, non-equilibrium currents are already present
even in the absence of friction.

\emph{General model}-- We consider a tracer of mass $M$ and velocity
$V$, the motion of which is constrained in one dimension, in contact
with small particles of one or more gases characterized by different
parameters: mass $m_i$, temperature $T_i$, density $\rho_i$,
restitution coefficient $r_i$ for the collisions between tracer and
gas molecules ($r_i\in [0,1]$, with $r_i=1$ for elastic collisions),
where the index $i$ denotes different gases.  Between successive
collisions with the gas particles the dynamics of the tracer is
affected by Coulomb friction, slowing its motion.  Furthermore, we
assume that the whole system, tracer plus gas(es), presents a spatial
asymmetry.

The dynamical evolution of the system is represented by a piecewise
deterministic process, described by stochastic jumps, modeling the
interactions of the tracer with the particles, and by a deterministic
term, taking into account the friction.  The differential equation
describing the probability density function of the tracer velocity
$P(V,t)$ is
\begin{eqnarray}
\frac{\partial P(V,t)}{\partial t}&=&\int dV'[W(V|V')P(V',t)-W(V'|V)P(V,t)]\nonumber \\
&+&\Delta\frac{\partial}{\partial V}\sigma(V)P(V,t),
\label{1.1}
\end{eqnarray}
where $\Delta$ is the frictional coefficient, $\sigma(x)$ is the sign
function and $W(V'|V)$ are the transition rates for the jump from $V$
to $V'$ due to collisions, which depend on the mass $M$ and the gas
parameters ($\rho_i$, $T_i$, $m_i$ and $r_i$).  The spatial asymmetry
may appear in the model through the following structure of the
transition rates
\begin{equation}
W(V'|V)= 
\begin{cases} W^+(V'|V) &  \textrm{if $V'>V$}
\\
W^-(V'|V) & \textrm{if $V'<V$,}
\end{cases}
\label{112}
\end{equation}
with $W^+(V'|V)\ne W^-(V'|V)$. This structure is realized for instance
by the probe with triangular shape considered
in~\cite{cleuren2,costantini1}, where collisions on one side always
accelerate the object whereas collisions on the other sides slow its
motion. The condition~(\ref{112}) is not sufficient for a ratchet
effect to be observed. In particular, if $\Delta=0$ and the transition
rates are assumed to satisfy the detailed balance (DB) relation with
respect to an equilibrium distribution $P_0(V)$, then, denoting by
$\langle\dots\rangle_0$ the average over $P_0(V)$, one has $\langle
V\rangle_0=0$ and $\langle V^2\rangle_0=T/M$, where $T$ is the
temperature of the thermal bath of particles (if several gases are
present, their temperatures have to be the same, i.e. $T_i=T$ for
every $i$).

We stress that model~(\ref{1.1}) is more general than models described
by Langevin equations (i.e. with white noise), widely studied in the
literature of Brownian motors~\cite{R02,HM03}. Indeed, in a Langevin
equation non-equilibrium conditions can only appear through
time-varying parameters (potentials or temperatures), or through
external forces. Here, the ``kinetic'' nature of noise is more
physical and, due to the different time scales involved, gives the
possibility to introduce non-equilibrium conditions such as constant
different temperatures or other dissipation channels.

In this model it is important to put in evidence the competition
between different timescales: the mean collision times which, in the
limit of large mass $M$ take the form $\tau_i\simeq
\sqrt{m_i/T_i}/(\rho_i S_i)$, with $S_i$ the scattering cross section
of the tracer with the particles of gas $i$, and the stopping time
$\tau_\Delta=V^*/\Delta$ due to friction, where $V^*$ is the average
velocity after a collision. These characteristic times introduce two
regimes in the dynamics. For $\min\{\tau_i\}\gg \tau_\Delta$, the
system is in the rare collision limit, where collisions always occur
when the tracer is at rest because of friction. In this case, the
dynamics evolves via slip-stick motions and the stationary
distribution develops a singular contribution in $V=0$, as explained
in~\cite{talbot1}.  In the opposite limit, when $\max\{\tau_i\}\ll
\tau_\Delta$, the system is in the frequent collision limit, and the
tracer is never at rest.

\emph{Ratchet effect and lack of equipartition}-- To obtain a general
expression for the average velocity of the tracer, we multiply by $V$
both members of Eq.~(\ref{1.1}) and integrate over $V$. In the
stationary state, we get for the momentum flow
\begin{equation}
0=-\Delta\langle\sigma(V)\rangle + \langle \alpha(V)\rangle,
\label{2.3}
\end{equation}
where $\alpha(V)=\int (V'-V) W(V'|V) dV'$ is the jump moment, which
depends on $M$ through the rates $W$, and the symbol
$\langle\dots\rangle$ denotes an average over the stationary
distribution $P(V)$.

For mass of the tracer large enough with respect to the largest mass
among those of the gas particles, denoted hereafter by $m$, we can do
an expansion around $V=0$~\cite{vK61}. Keeping terms up to the second
order, we obtain
\begin{equation}
\alpha+\alpha'\langle V\rangle+\frac{1}{2}\alpha''\langle
V^2\rangle-\Delta\langle \sigma(V)\rangle\simeq 0,
\label{2.4}
\end{equation}
where $\alpha'$ and $\alpha''$ denote the first and second derivatives
of $\alpha$ with respect to $V$, respectively, and all coefficients
are computed in $V=0$.  In particular, $|\alpha'|^{-1}$ represents the
characteristic thermalization time $\tau_{th}$ of the tracer with the
gas, in the absence of friction. This time scale is related to the
collision time: $\tau_{th}\sim \tau_i M/m_i$~\cite{vK61}. The
coefficients $\alpha$, $\alpha'$ and $\alpha''$ are functions of $M$
through the transition rates, and have to be expanded in powers of
$M^{-1}$ consistently, taking into account that $\langle V^2\rangle
\sim \mathcal{O}(M^{-1})$.  Eq.~(\ref{2.4}) yields
\begin{eqnarray}
\langle V\rangle &=&-\frac{1}{\alpha'}\left[\alpha+\frac{1}{2}\alpha''\langle V^2\rangle\right]
+ \frac{\Delta}{\alpha'}\langle \sigma(V)\rangle \nonumber \\ 
&=&-\frac{A}{\alpha'}\left[T_k-T \right]+\frac{\Delta}{\alpha'}\langle \sigma(V)\rangle,  
\label{2.6}
\end{eqnarray}
where in the second line we have assumed that thermal gradients (if
present) are small so that one can define a base temperature $T$, and
we have introduced the kinetic temperature $T_k\equiv M\langle
V^2\rangle$ (assuming $\langle V\rangle^2\ll \langle V^2\rangle$) and
a general asymmetry $A$ through the expressions
\begin{equation}
\alpha\simeq-T A, \qquad \alpha''\langle V^2\rangle\simeq 2AT_k.
\label{2.7}
\end{equation}
The above structure for the coefficients $\alpha$ and $\alpha''$ is
verified in many
examples~\cite{meurs1,vandenbroeck,cleuren2,costantini2,cleuren}
including the one discussed below, and follows from
Eq.~(\ref{112})~\cite{note} (the explicit form of $A$ depends on the
specific model).

The interest of Eq.~(\ref{2.6}) is in making clear that there are two
contributions to the ratchet's drift, corresponding to the two
channels for heat exchanges of the probe: the first one is
$D_{hf}\equiv-\frac{A}{\alpha'}\left[T_k-T \right]$, which is
proportional to the temperature difference $T_k-T$, and therefore to
the heat flux exchanged between the ratchet and the thermal bath,
induced by collisions; the second one is directly related to the
presence of friction and is proportional to the average of the
frictional force: $D_\Delta\equiv \Delta\frac{M}{\alpha'}\langle
\sigma(V)\rangle$. Notice that the first channel can be sustained by
the presence of \emph{reservoirs} at different temperatures or
dissipative collisions, but it is also affected by the presence of
friction. Indeed, if elastic interactions are considered and all the
baths are at the same temperature, a net flow can still be generated
by frictional dissipation.

For $\Delta=0$, or when the thermalization time $\tau_{th} \sim
1/|\alpha'|$ is small with respect to the stopping time $\tau_\Delta$
and friction can be neglected, only the term $D_{hf}$ remains. Then
the ratchet effect can be present if and only if the transition rates
are asymmetric (i.e. $A\ne 0$) and do not satisfy DB, so that the
kinetic temperature $T_k$ is different from that of the external bath
$T$. This is the case for many collisional ratchets studied in the
literature~\cite{meurs1,vandenbroeck,cleuren2,costantini2,cleuren},
where the explicit expressions obtained for the drift in the different
cases can be put in a form analogous to the first term of
Eq.~(\ref{2.6}).  In the opposite regime, namely when friction
dominates, both channels are active and the two contributions may
produce interesting interplays, with non-monotonic behaviors in the
drift, as shown below. An expression reproducing the non-monotonic
drift for collisional ratchets (which is a feature already observed,
e.g. in~\cite{gnoli}) represents a relevant result of the present
study.

\begin{figure}[!t]
\includegraphics[width=1.\columnwidth,clip=true]{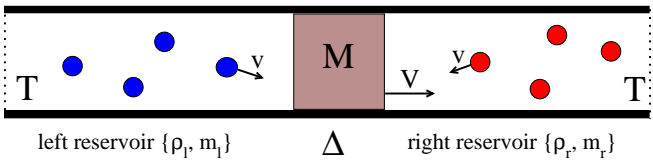}
\caption{The asymmetric Rayleigh piston with Coulomb friction.}
\label{fig0}
\end{figure}

\emph{The asymmetric Rayleigh piston}-- To make more explicit our
discussion, we now consider the generalized Rayleigh
piston~\cite{macdonald,PS04}, in the presence of Coulomb friction.  It
consists of a piston of mass $M$, the two faces of which are connected
with two different gases of elastic particles of mass $m_r$ (at right)
and $m_l$ (at left), see Fig.~\ref{fig0}. The two gases are at
equilibrium at the same temperature $T$ and have densities
$\rho_i=\rho$, with $i=r, l$. In such a way the pressures on both
sides of the piston are equal. The piston velocity is changed by the
elastic collisions with the (right and left) gas particles according
to the rule $V'=V+\frac{2}{1+M/m_i}(v-V)$, where $V$ and $V'$ are the
piston velocities before and after the collision, respectively.  The
particles velocities are distributed according to the
Maxwell-Boltzmann distribution $p_i(v)=\rho_i\sqrt{\frac{m_i}{2\pi
    T}}\exp\left(-\frac{m_iv^2}{2T}\right)$, where the Boltzmann's
constant $k_B=1$.

\begin{figure}[!t]
\includegraphics[width=0.9\columnwidth,clip=true]{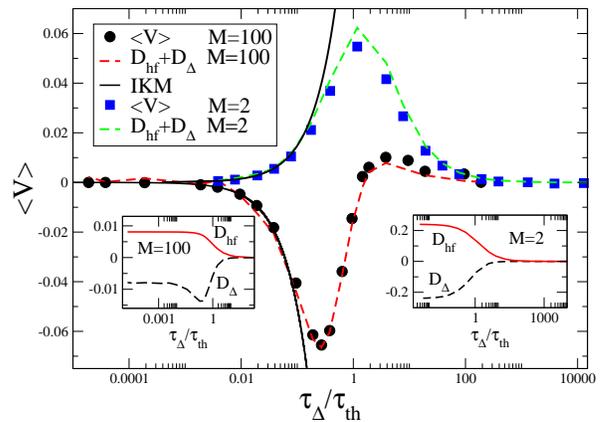}
\caption{(color online) DSMC results for $\langle V\rangle$, with
  $T=10$, $\rho=0.5$, $m_r=2$ and $m_l=1$, as a function of
  $\tau_\Delta/\tau_{th}$, for $M=100$ (black dots) and $M=2$ (blue
  squares). The black curves show the analytical predictions of the
  IKM, Eq.~(\ref{drift}). The dashed lines (red for $M=100$ and green
  for $M=2$) represent the rhs of Eq.~(\ref{2.6}) with $\langle
  V^2\rangle$ and $\langle\sigma(V)\rangle$ computed in DSMC. The
  amplitude of $\langle V\rangle$ for $M=100$ has been magnified by a
  factor 10, for scale reason. Insets: $D_{hf}$ (red curves) and
  $D_\Delta$ (black dashed curves) \emph{vs} $\tau_\Delta/\tau_{th}$.}
\label{fig1}
\end{figure}

In this model the asymmetric transition rates are~\cite{macdonald}
\begin{eqnarray}
W^+(V'|V) &=& \left(\frac{M+m_l}{2m_l}\right)^2(V'-V) \nonumber  \\
&\times& p_l\left(\frac{M+m_l}{2m_l}V'-\frac{M-m_l}{2m_l}V\right),  \nonumber \\
W^-(V'|V) &=& \left(\frac{M+m_r}{2m_r}\right)^2(V-V') \nonumber \\
&\times& p_r\left(\frac{M+m_r}{2m_r}V'-\frac{M-m_r}{2m_r}V\right).  
\label{ratep}
\end{eqnarray}
These transition rates satisfy DB with respect to the Gaussian
distribution $P_0(V)=(2\pi T/M)^{-1/2}\exp(-MV^2/2T)$.  The explicit
expressions for the coefficients appearing in Eq.~(\ref{2.6})
are~\cite{macdonald}:
\begin{eqnarray}
\alpha&=&\rho T[(M+m_l)^{-1}-(M+m_r)^{-1}] \nonumber \\
&=&-\rho T(m_l-m_r)/M^2 + \mathcal{O}(M^{-3}), \label{alpha1} \\
\alpha'&=&-2\rho \sqrt{\frac{2T}{\pi}}\left[\frac{\sqrt{m_l}}{M+m_l}+\frac{\sqrt{m_r}}{M+m_r}\right], \label{alpha1p} \\
\alpha''&=& 2\rho [m_l/(1 + m_l/M) - m_r/(1 + m_r/M)]/M \nonumber \\
&=&2\rho(m_l-m_r)/M+\mathcal{O}(M^{-2}).
\label{alpha1s}
\end{eqnarray}
From Eqs.~(\ref{alpha1}) and~(\ref{alpha1s}) follows that the explicit
formula for the asymmetry is $A\simeq\rho(m_l-m_r)/M^2$, which
justifies the relations~(\ref{2.7}). A similar structure for the
coefficients $\alpha$ and $\alpha''$ can be traced back in many
collisional
ratchets~\cite{meurs1,vandenbroeck,cleuren2,costantini2,cleuren}.  In
this model the time scales are
$\tau_\Delta=V^*/\Delta=\sqrt{T/M}/\Delta$, because collisions are
elastic, and $\tau_{th}=1/|\alpha'|\simeq
\sqrt{\pi/(2T)}M/[2\rho(\sqrt{m_l}+\sqrt{m_r})]$.

To study the behavior of the model and to verify the
relation~(\ref{2.6}) in all regimes, we perform numerical simulations
of the process~(\ref{1.1}) with transition rates~(\ref{ratep}), using
a Direct Simulation Monte Carlo (DSMC) algorithm~\cite{B94}. We
extract the velocity $v$ of a gas particle from $p_i(v)$, $i=r, l$
with probability 1/2, and then we allow the collision with the piston
with velocity $V$ to occur with probability $\propto|v-V|$. In
Fig.~\ref{fig1} $\langle V\rangle$ is shown (black dots for $M=100$
and blue squares for $M=2$) as a function of the ratio
$\tau_\Delta/\tau_{th}$, which is varied by changing $\Delta$, with
the other parameters fixed (see caption). A net drift is found in a
wide range of $\Delta$ values: we stress that, at variance with
kinetic models studied previously, in this system the ratchet effect
is entirely driven by the Coulomb friction, because the two gases are
in equilibrium at the same temperature and collisions are elastic.

The complex non-monotonic behavior of the drift is very well described
in all the regimes by the r.h.s. of Eq.~(\ref{2.6}), represented in
Fig.~\ref{fig1} by red (for $M=100$) and green (for $M=2$) dashed
lines. The parameters $\alpha, \alpha'$ and $\alpha''$ are given by
Eqs.~(\ref{alpha1}),~(\ref{alpha1p}) and~(\ref{alpha1s}) and the
averages $\langle V^2\rangle$ and $\langle\sigma(V)\rangle$ are
computed in DSMC.  The behavior of the two terms $D_{hf}$ and
$D_\Delta$ is reported in the insets of Fig.~\ref{fig1}.  Both terms
display plateaux in both the opposite limits $\tau_\Delta \ll
\tau_{th}$ and $\tau_\Delta \gg \tau_{th}$. The plateau in the latter
limit is zero for both terms, as equilibration with the thermal bath
is quickly attained, inducing a zero drift.  Since also in the
opposite limit of rare collisions the drift is expected to vanish, as
shown below, this produces the peaks observed in
Fig.~\ref{fig1}. Notice that in this model Eq.~(\ref{2.6}) also holds
for values of $M$ comparable to those of the gas particles (see
Fig.~\ref{fig1}), if all orders in $M$ are retained in
expressions~(\ref{alpha1}),~(\ref{alpha1p}) and~(\ref{alpha1s}). This
is due to the specific forms of the coefficients: in particular, all
even derivatives of $\alpha(V)$ greater than the second one vanish for
this model~\cite{macdonald}.

\emph{Independent kick model}-- An analytical explicit formula for the
average drift can be obtained in the physical situation of rare
collisions, namely when $\min\{\tau_i\}\gg\tau_\Delta$. In this case,
assuming that every collision occurs when the piston is at rest, the
average velocity can be computed in the so-called Independent Kick
Model (IKM)~\cite{talbot1,talbot2}.  For our model this yields
\begin{equation}
\langle V\rangle = \left(\int dv |v|p_r(v)+\int dv |v|p_l(v)\right)\int_0^\tau V(t)dt,
\end{equation}
where $V(t)=V_0-\Delta\sigma(V_0) t$, $\tau=|V_0|/\Delta$ and $V_0$ is
the velocity after a collision: $V_0=V^+$ if $v>0$, and $V_0=V^-$ if
$v<0$, where $V^+=\frac{2v}{1+M/m_r}$ and $V^-=\frac{2v}{1+M/m_l}$.
Using these expressions one obtains
\begin{equation}\label{drift}
\langle V\rangle=\frac{2\rho}{\Delta}\sqrt{\frac{2T^3}{\pi}}
\left[\frac{\sqrt{m_l}}{(m_l+M)^2}
-\frac{\sqrt{m_r}}{(m_r+M)^2}\right].
\end{equation}
In this formula the net drift explicitly appears when the asymmetry in
the system is present (i.e. $m_r\ne m_l$).  For small $\Delta$ the
formula is not expected to hold because the approximation of rare
collisions is not valid. Notice also that in the limit $M\to\infty$
the drift vanishes.  In Fig.~\ref{fig1}, the analytical
prediction~(\ref{drift}) of the IKM (black lines) is shown to be in
perfect agreement with the numerical results in the rare collision
regime. Fig.~\ref{fig1} also shows that formula~(\ref{2.6}) is in
agreement with the IKM prediction.

\emph{Conclusions}-- We have presented two interesting results: i)
formula~(\ref{2.6}) for the average drift of a general collision
ratchet in the presence of friction can describe the ratchet behavior
in all regimes, and explicitly shows the two channels of dissipation
contributing to the drift; this relation has been also tested in a
rotor ratchet with dry friction recently studied in~\cite{gnoli}; ii)
Coulomb friction can be a source of dissipation sufficient to generate
a ratchet effect in thermal baths. Our study can be extended to other
forms of non-linear friction~\cite{UKGI04,PF07}.

Our results on the ratchet effect driven by Coulomb friction in a
thermal bath pave the way to applications in the field of
nanophysics. At these scales, thermal fluctuations can be induced by a
gas of molecules or a liquid environment, and the Coulomb friction is
still present, as well known from atomic friction
experiments~\cite{GTVT10}. Moreover, the developments of new
techniques for the design and fabrication of nano-devices can provide
probes with desired shapes and asymmetries. Therefore, all the
ingredients are available to realize ratchet devices at small scales
entirely based on the action of Coulomb friction as source of
dissipation.

\emph{Acknowledgments}-- The work of the authors is supported by the
``Granular-Chaos'' project, funded by the Italian MIUR under the
FIRB-IDEAS grant number RBID08Z9JE.

\bibliography{fluct}

\begin{thebibliography}{10}
\bibitem{note}%
  \BibitemOpen
  \bibinfo {note} {For models where the probe is in contact with baths at
  different temperatures, $T$ is the average of temperatures weighed by the
  masses of the gases.}%
  \bibAnnoteFile{Stop}{note}%
\end{thebibliography}

 \end{document}